Report on the ESO workshop

# VLT Beyond 2030 and Call for White Papers

held at ESO Headquarters, Garching, Germany, 26–30 January 2026

Céline Péroux[1]
Antoine Mérand[1]
Martyna Chruślińska[1]

[1] ESO

The VLT Beyond 2030 conference gathered participants from the community and ESO experts to present and discuss science and technological ideas for the future of both the Very Large Telescope (VLT) and its interferometer (VLTI). An effort was made to pair participants with science and engineering backgrounds so as to optimise cross-field fertilisation and initiate 'out-of-the-main-room' discussions. The well-attended conference reflected the continued interest within our community in developing new projects for the VLT/I and the will to use this facility to answer key science questions that are likely to be crucial in the decades to come. This article presents a summary of the overall points of focus during the conference and also serves as the opening of the call for White Papers with a deadline of 15 January 2027. Proposers are encouraged to fast-forward to the years beyond 2030 to identify essential areas of research so that together we can shape a long-term roadmap for the VLT/I.

## Motivation

ESO operates optical/near-infrared telescopes on its La Silla and Paranal observatory sites and is a partner in the Atacama Large Millimeter/submillimeter Array (ALMA). The Extremely Large Telescope (ELT) on Cerro Armazones is under construction, expected to start operations at the end of the decade. ESO is a partner in the Cherenkov Telescope Array (CTA) which will start operations in the coming years. In addition, ESO provides resources such as user support and science archives to empower the community to make the best use of the facilities offered. The existing and planned ESO facilities cover a large parameter space across the electromagnetic spectrum, angular, spectral and time resolution, multiplexing and observing methods (Brinchmann et al., 2025a). In particular, Paranal observatory was developed specifically to host the Very Large Telescope (VLT) and its interferometer (VLTI) some 30 years ago. The VLT/I instrumentation has been developed over many years and continues to benefit from developments and upgrades.

The VLT Beyond 2030 process aims to maintain the VLT/I at the forefront of astrophysical research in the decades to come. This is distinct from the Expanding Horizons[1] process (Brinchmann et al., 2025b), the goal of which is to select the next ESO facility after the ELT. The development of the VLT/I instruments follows planning that is discussed with the ESO governing bodies to derive the best synergies amongst facilities. As part of this process, ESO organised a conference, named VLT Beyond 2030, at its headquarters in Garching near Munich (Germany) in January 2026. The programme[2] covered many topics and included extensive discussion sessions to promote ideas and thought-provoking exchanges beyond the invited and contributed contributions. This article provides a short, non-exhaustive summary of the meeting and also serves as the opening of the call for White Papers.

## Bird's eye view

At the start of the conference, the ESO senior management reminded everyone of what a privilege it is to be able to observe, and hence our duty to protect our skies from multiple types of pollution from the ground, including light, dust and vibration. Later in the conference, light pollution from space, as generated by, for example, satellite constellations, was also discussed. While recent events have been positive in this respect[3], the astronomical community will have to keep high on its agenda an effective and coordinated dialogue with stakeholders to preserve the purity of the sky for the good of astronomical research and, more broadly, of human knowledge.

The start of the conference also included various presentations on the impact of future VLT/I instrumentation on sustainability. As our society is facing major environmental challenges, future projects will have to monitor and minimise their environmental impact. In addition, the consortia building the instrument beyond 2030 are invited to encourage diversity in their teams. While the timescale envisioned naturally provides opportunities for early-career scientists to contribute, a good mix of gender and minority representation would be an obvious asset for future projects.

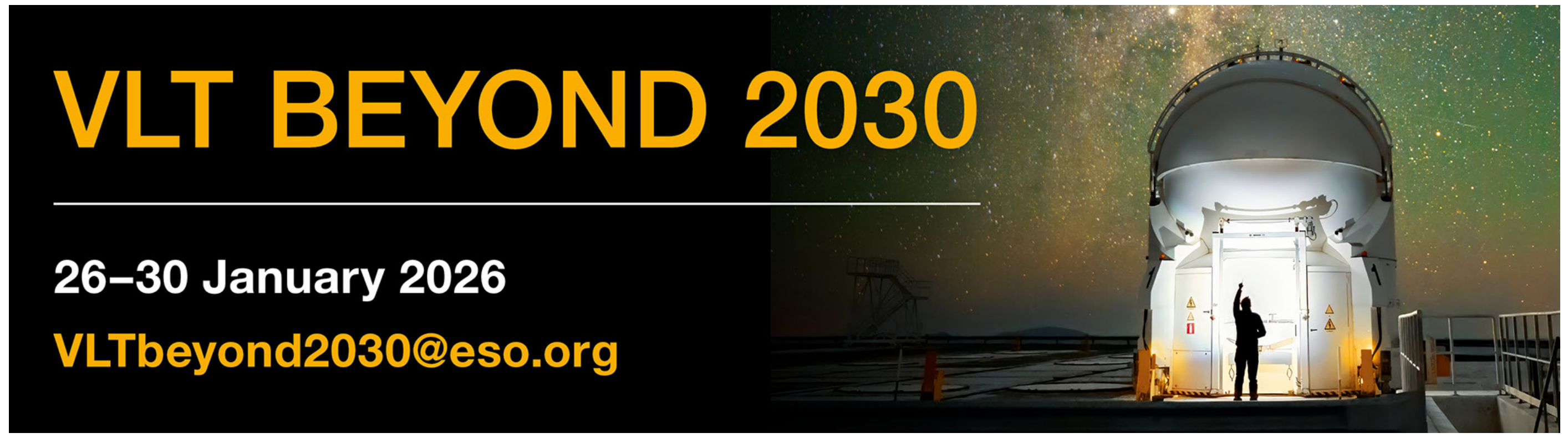

Figure 1. VLT/I Beyond 2030 workshop banner.

The first day of the conference provided overviews of both ground- and space-based facilities at multiple wavelengths. It is becoming obvious that a number of future missions’ science goals cannot be fulfilled without coordinated follow-up from the ground. A discussion session dedicated to these topics concluded that an assessment of the ground-based segment as early as at the proposal stage, perhaps by having representatives of the partner agency taking part in the review process, would be the best way to coordinate such essential efforts efficiently. The role of public surveys in that context was also highlighted. Ultimately, this coordination would greatly benefit from synchronised roadmapping exercises, which would take into account similar timescales and scope.

All week, speakers stressed the several aspects unique to the VLT/I which should be not only preserved but also further developed. This included coherent and intensity interferometry and a blue coverage which remains competitive with respect to the ELT thanks to both a three-mirror design and the chosen mirror coatings. Similarly, the availability of four Unit Telescopes (UTs) makes the VLT/I unique in terms of survey power with respect to the next generation of giant telescopes, perhaps providing a means to select targets to then follow up with the ELT. It was highlighted that gathering statistics is going to be key for cosmology-related research in particular. Importantly, the VLT/I and even the Auxiliary Telescopes provide a valuable testbed for new technological developments and on-sky tests. Discussions included the idea of also awarding technological-development time (in addition to science time) on the VLT/I and driving new discovery by enabling technological breakthroughs. Some of these could be done via the recent new possibility of requesting Guaranteed Time Observations as part of an ESO Tech Dev project.

Some of the conference discussions highlighted that part of the future needs are at the level of the facility itself. Ideas included equipping all UTs with multiple laser guide stars and deformable mirrors to enable the use of adaptive optics (AO) for most future instruments while lowering the pressure on UT4. Furthermore, there were suggestions to provide a clear roadmap on the VLT/I instrumentation plan in the next decades to ensure scientifically valuable monitoring programmes can be sustained without instruments being unexpectedly decommissioned. Indeed, an obsolescence programme which preserves the efficient operation and maintenance of both successful current instruments and the telescopes themselves appears important. Finally, questions on how rapidly emerging private observatories might be integrated in the current system were also debated.

## Scientific perspectives

The science landscape in the 2030s and beyond is paved with many challenging questions. Several conference attendees stressed that catering for multimessenger astronomy, especially as we stand on the verge of an era of booming numbers of transient triggers, will become increasingly important. Multimessenger astronomy beyond the 2030s will also require rapid access to real-time and archival data from multiple observatories to rapidly identify and follow up on crucial targets. This may necessitate a rethink of how we communicate alerts and design databases, a field in which ESO could play a coordinating role. As the number of astronomical objects classified as transients increases, the need for spectroscopic observations across the electromagnetic spectrum becomes more obvious and public surveys or alternatives might play a role in coordinating the community requests. In synergy with multiple other facilities targeting new messengers, including very high-energy photons, neutrinos or gravitational waves, the VLT/I is bound to have a role to play in these areas of research.

Cosmologists are still pursuing an understanding of the nature of dark energy and dark matter, while the so-called Hubble tension remains an uncomfortable challenge to our understanding of the largest scales of the Universe. Big Bang nucleosynthesis and related topics, such as the cosmic lithium problem, helium-4 abundance and deuterium over hydrogen abundance, have seen progress in recent years. High on the agenda is a measurement of the mass of neutrinos for which progress will likely require synergy with other experiments. Possible contributions by the VLT to these topics in the years to come include measurements of the variation of fundamental physical ‘constants’ at a level predicted by some theories and the so-called Sandage test measuring the expansion of the Universe through redshift drift. In the field of galaxy formation, a refined understanding and characterisation of the baryon cycle which connects the intergalactic medium with galaxies (dubbed intragalactic science) and stellar- and black hole-driven feedback processes will likely be key in 2030 and beyond. Some of the current VLT instruments have recently contributed crucial results to these fields and more are expected in the decades to come. The physics of black holes, especially the massive objects observed at high redshift, is another important area of development, with both ALMA and the VLT providing important spectroscopic results. The search for the elusive Population III stars is another active research domain. The epoch of reionisation and the fraction of ionising photons escaping galaxies also represent a dynamic field of research.

In the booming field of exoplanets, the role of spectroscopy was emphasised, highlighting that planet detection relies critically on long time series. A rapidly expanding sub-field is aiming to characterise the hot and cold atmospheres of exoplanets as a key component of both their chemical abundance and dynamics. It was stressed that a mono-mode fibre instrument at red wavelengths would enable us to search for habitability in Earth-mass planets orbiting late M-dwarfs. On the other hand, high-contrast imaging enabled by coronagraphy, coupled with AO and advanced software post-processing, offers the prospect of effective direct imaging approaching the fundamental limit. In particular, machine learning and predictive control of the AO systems provide a promising avenue for the future. Despite the limitations related to telescope availability and timescale for tailored instruments, the more than two orders of magnitude gain from the ELT with respect to the VLT in terms of direct imaging was stressed. As for other fields, the uniqueness of the VLT at shorter wavelengths remains. Exoplanet spectroscopy and astrometry would also remain competitive with the VLTI, by expanding to shorter wavelengths and higher spectral resolution. Additionally, integral field spectroscopy in reflected light at the VLT could harvest potential biosignatures of the most promising targets. Finally, an improved understanding of planetary systems across time and space will be an important component of research in the field.

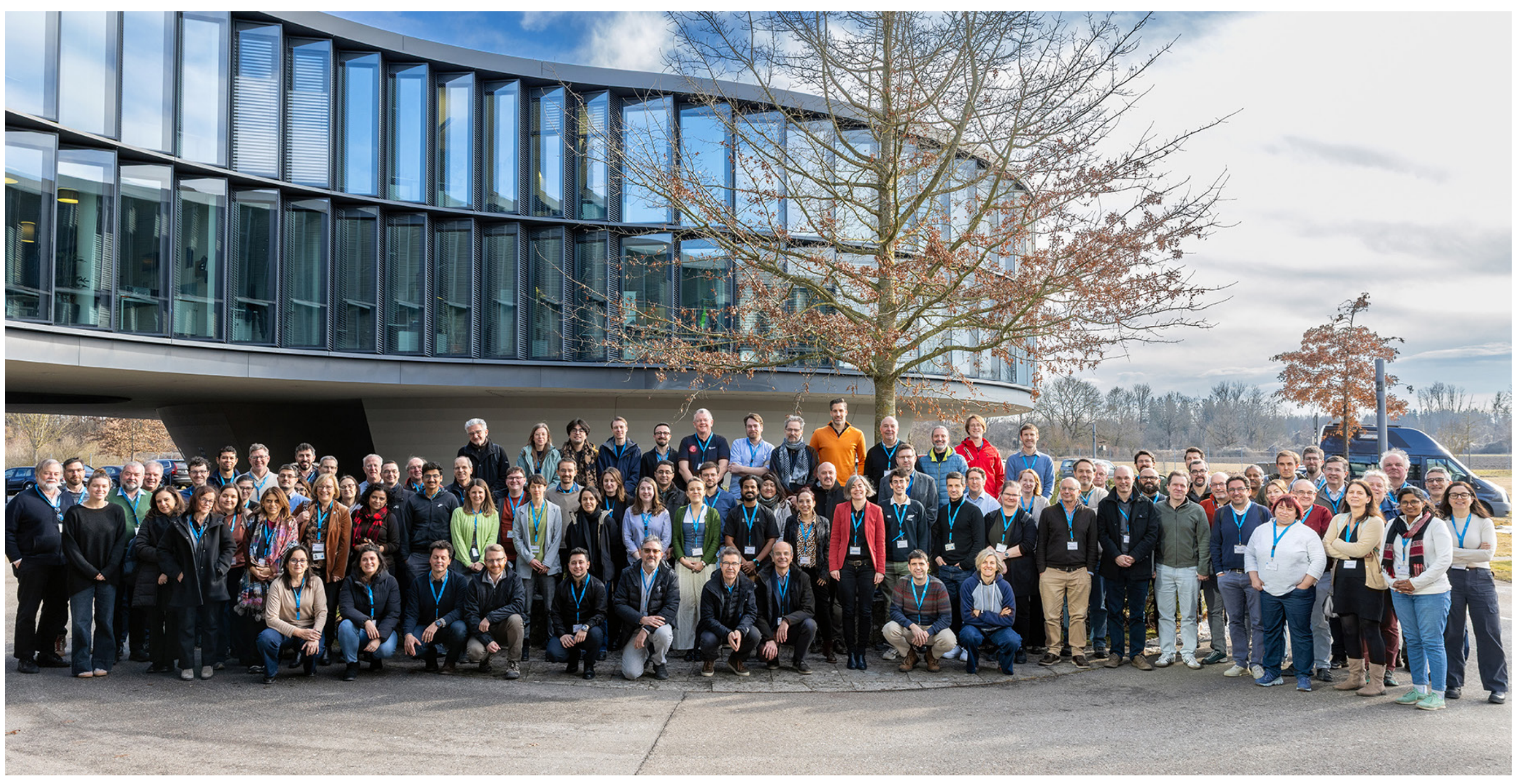

Figure 2. Many participants braved the harsh German winter to take part in the conference.

On the massive stars front, many questions will be addressed in the coming years including: i) probing the formation and evolution of binaries and other multiples; ii) quantifying the effect of metallicity; and iii) characterising stellar feedback. Open questions about star formation include population studies, accretion and ejection processes, disc physics and determining the universality of the initial mass function. Likewise, stellar structure, as probed through astroseismology and including magnetospheres, will likely be an area of development alongside studies of stellar remnants. Equally important is work on the origin of elements and nucleosynthesis as well as hierarchical galaxy assembly — including testing the lambda-cold-dark-matter scenario outside our Milky Way, but also establishing the number of merging events and nucleocosmochronology using chemical abundances to measure the ages of the oldest stars.

Other topics not developed in detail at the conference include the Solar System and its giant planets, also comprising comets and other fast-moving objects, as well as so-called galactic archeology.

## Parameter space

One way to envision the next generation of instruments is naturally to expand the parameter space which will then likely enhance the discovery potential. Natural parameters to advance include i) spectral resolution, ii) angular resolution, iii) astrometic accuracy, iv) wavelength coverage, v) field of view, vi) multiplexing, vii) time-domain, viii) wavelength accuracy, ix) optimised transmission and x) contrast. Building on highly successful current instruments, pushing single-object, high-resolution spectroscopy to redder wavelengths on both the VLT and La Silla telescopes was proposed in various forms by several contributors. Likewise, current multi-object spectrographs cover a wide range of scientific topics from galactic archeology to the high-energy sky, galaxy evolution and even cosmology. The need for very high-resolution multi-object spectroscopy, an ESO legacy, was raised on several occasions, stressing in particular the role it might play in following up past & future European space missions. In the last decades, large-field integral field spectrographs (IFUs) have transitioned from single-object instruments to panoramic facilities catering for a large variety of science fields by providing massive spectroscopy in dense fields, both for extended sources and in deep fields. As we are entering an era of wide and deep photometric surveys (with facilities like Euclid, Vera C. Rubin Observatory and the Nancy Grace Roman space telescope), the need for matching spectroscopic surveys becomes even more obvious. Transmission remains a key element of such an instrument, in turn requiring optimisation of mirror coatings as well as further developments in the area of volume-phase holographics. Moving to redder wavelengths with such a wide-field instrument would offer a new parameter space. The challenges will be to cut the cost per spectrograph while providing a compact, and hence a more manageable, design. It was argued that, simply put, IFUs provide “spectroscopy of everything”.

In addition to these, other boundaries of the parameter space were proposed during the week. Various conference participants emphasised that line fidelity (which is different from wavelength accuracy) and simultaneity in terms of wavelength and also continuous spatial coverage are areas of potential advances. Additionally, brokers to triage the interesting targets before triggering a rapid response mode and appropriate data flow to enable quick access to the data are key aspects of the parameter space. Speakers also highlighted the power of low-surface-brightness sensitivity of extended sources as a tool to challenge the lambda-cold-dark-matter model. Polarimetry at various wavelengths, where multiple proofs of concept already exist, was felt to be an important tool to improve on our physical understanding of a wide number of astrophysical sources, ranging from planets, stars and local galaxies to supernovae and active galactic nuclei.

## Key technological developments

From its inception, the conference aimed at associating scientific prospects beyond 2030 with innovative technologies at various levels of maturity. The science organisation committee suggested pairs of reviewers with either scientific or technological expertise on topics which might mutually benefit each other. This setup naturally highlighted a number of important current and future developments which will likely turn out to be essential for the VLT/I beyond 2030. Work on detectors and controllers is an obvious area of development, including the potential for low-noise, high quantum efficiency and curved detectors to enable instruments with larger fields of view. Maintaining supply chains of various sub-systems (detectors, thorium hollow-cathode lamps, etc) was also felt to be important. A discussion session reported a need for an increase in the maturity of accurate wavelength calibration from both (pressure-controlled) hollow-cathode lamps and laser frequency combs to new technologies, including electro-optic modulators and silicon chips. Panelists also stressed the challenges of interfacing these calibration units with the instrument/telescope. It was pointed out that for exoplanet research a laser frequency comb was not needed every night so that one system could be used by several telescopes and/or UTs. Reports of using a filter-tilter in front of the telescope optics to enable narrow-band filtering without degrading the flux transmission were also provided. Equally important are future developments in the field of AO, including tip-tilt correction to enable full-sky AO, more powerful laser guide stars enabling laser splitting, or orbiting artificial stars offering brighter sources with no cone effects and no need for stabilisation, as well as stable deformable mirrors. Intensity interferometry offers the potential to resolve 50-microarcsecond scales by combining the VLT and the 4-metre Visible and Infrared Survey Telescope for Astronomy (VISTA) telescope in Paranal, while robotic fibre positioners have been shown to be smaller and faster and to offer more accurate positioning, allowing for high-fibre-density instrument design. Virtually imaged phased array dispersers might enable a path towards ultra-high ($R > 300\,000$) spectral resolution, while the potential of astrophotonics, including for OH line suppression, has been highlighted. Clearly, more compact instruments would offer the benefits of lower costs, requiring fewer resources and lighter associated logistics.

## Thinking outside the box

Beyond these technological developments, several participants emphasised the importance of adapting the VLT/I operation model to new needs, including high flexibility. This encompasses rapid follow-up of the many transients soon to be available, monitoring targets over several years and enabling a route for possible high-gain, high-risk proposals.

It was stressed several times that, beyond hardware, successful science projects also benefited from innovative software. While data reduction software at ESO has reached a high level of maturity, the instruments to be operated in the next decade likely will have to provide equally robust data analysis software to maximise the science outputs in a multi-wavelength environment. It was stated that this is particularly true for upcoming panoramic IFUs for which essentially every observation is a survey. The ever-increasing value of an easy-access, interoperable and public archive was also stressed. In an era of large data sets, the role of machine learning based on foundation models to harvest such archives is likely to become increasingly relevant. Importantly, it was highlighted that some of the astrophysical measurements to be made beyond 2030 are limited by flux calibration, highlighting the need for better spectrophotometric featureless standard stars and also for improvements to poorly constrained laboratory measurements of atomic data, which prove nonetheless key to deriving the composition and abundance of various elements. Research in these areas needs to be supported and promoted to enable the best scientific output.

## Conference demographics

The Scientific Organising Committee (SOC) sought fair representation from the community. The SOC aimed for a balanced ratio of female to male speakers for invited and contributed talks compared to the ratio among participants. As a result, the SOC itself was composed of 5/6 F/M (45% female), while the invited speakers were 6/13 F/M (46% female) and the session chairs 10/22 F/M (45% female). The proposed contributed talks (79 total) were split 16/59/4 F/M/”prefer not to say” (20% female) while the actual contributed speakers (35 total) were 9/24/2 F/M/”prefer not to say” (25% female).

Attendees came from six continents (all but Antarctica), with the following percentages: 83.4 % Europe (Germany, Italy, France, Netherlands, Switzerland, Austria, UK, Sweden, Spain, Portugal, Denmark, Ireland, Belgium [r], Czechia [r], Poland [r], Finland [r], Hungary [r]); 6.3 % South America (Chile, Brazil); 4.9 % Australia; 2.4 % Asia (India, Iran [r], Turkey); 1.5 % North America (US, Canada) and 1.5% Other (where r = remote participants only).

The conference had a high oversubscription rate and attendance was over 140 in-person participants and typically 50 on-line attendees[4].

## Call for White Papers

As part of the VLT Beyond 2030 process, ESO is issuing a call for White Papers for new instruments for the VLT/I, with a deadline of 15 January 2027. The process aims to keep the VLT/I at the forefront of astrophysical research in the coming decades. The White Papers will be assessed by ESO and community representatives. The assessment phase will take place during 2027, leading to a phase A which will start in 2028 at the earliest. ESO will contribute hardware budget up to 10 million Euros per instrument.

A web page[5] provides detailed information about the format of the White Papers. Two separate documents need to be prepared: one focusing on science and the second one providing further technical information. The second document shall include specifically the following sections: i) instrument design elements; ii) technological developments; iii) consortium structure; iv) a management plan including a risk assessment; v) a budget plan; and vi) impact (societal, environmental, etc.) Together the two documents (in pdf format) should have a maximum of 50 pages, including figures but excluding references. The number of pages split between the two documents is up to the proposers. Submission instructions for these materials will be provided on the dedicated webpage[5].

All are invited to submit a White Paper, regardless of whether they have attended the conference or not. Projects at different levels of maturity are welcome as ESO aims to use this material to establish a roadmap for the VLT/I beyond 2030.

## Conclusion

Clearly, the week-long conference provided a forum in which to forge new connections between various experts, with the hope that these initial conversations will develop in well-shaped ideas by the deadline for the White Papers. The flexible format of the call is meant to encourage ambitious and transformational project proposals at various stages of maturity as ESO aims at using this material to establish a roadmap for the VLT/I beyond 2030. Ultimately, the VLT/I Beyond 2030 process aims to keep the VLT/I at the forefront of astrophysical research in the coming decades.

### Acknowledgements

We would like to thank the SOC for shaping an exciting conference programme. We are further grateful to the Local Organising Committee for volunteering their expert support to the organisation of the conference. Special thanks go to Denisa Tako for professional help and steadily good mood.

### Links

[1] Expanding Horizons: https://next.eso.org
[2] VLT Beyond 2030 programme: https://www.eso.org/sci/meetings/2026/VLT_beyond_2030/programme.html
[3] Plans for industrial complex near Paranal cancelled: https://www.eso.org/public/news/eso2602/
[4] Conference participants: https://www.eso.org/sci/meetings/2026/VLT_beyond_2030/list_participants.html
[5] Call for White Papers: https://www.eso.org/sci/meetings/2026/VLT_beyond_2030/whitepaper_call_vltbeyond2030.html

L. Sbordone/ESO

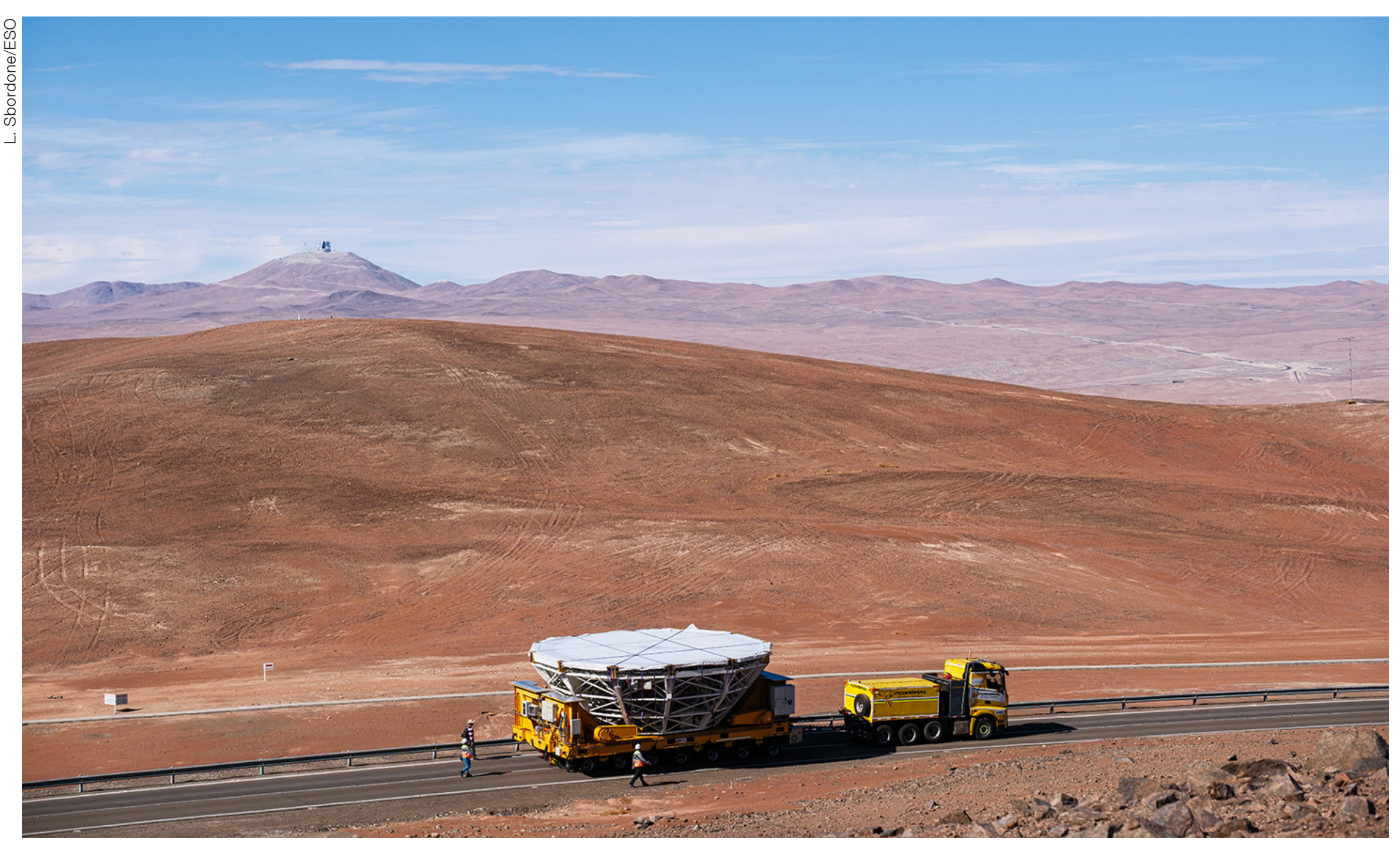

Have you ever wondered how a telescope keeps its mirrors in the best condition to observe the cosmos? In this image, a truck carefully carries one of the mirrors of ESO's Very Large Telescope (VLT), wrapped to protect it from the harsh environment of Chile's Atacama Desert. Its destination is the recoating facility that keeps the mirrors of this telescope perfectly shiny.